\documentclass[aps,prc,preprint]{revtex4}
 \usepackage{amsfonts}
\usepackage{amssymb}
\usepackage{graphics}
\usepackage{graphicx}
\begin{document}

\title{Specific heat capacity in the low density regime of asymmetric nuclear matter}

\author{R. Aguirre}
\email{aguirre@fisica.unlp.edu.ar}
\affiliation{Departamento de Fisica, Facultad de Ciencias Exactas, \\Universidad Nacional de La Plata, \\
and IFLP, CCT-La Plata, CONICET. Argentina.}

\begin{abstract}
Thermal and isospin composition effects over the heat capacity of
infinite nuclear matter are studied within the binodal coexistence
region of the nuclear phase diagram. Assuming the independent
conservation of both proton and neutron densities, a second order
phase transition is expected, leading to a discontinuous behavior
of the heat capacity. This discontinuity is analyzed for the full
range of the thermodynamical variables coherent with the
equilibrium coexistence of phases. Two different effective models
of the nuclear interaction are examined in the mean field
approximation, the non-relativistic Skyrme force and the covariant
QHD formulation. We found qualitative agreement between both
descriptions. The discontinuity in the specific heat per particle
is finite and decreases with both the density of particles and the
isospin asymmetry.\\
As a byproduct the latent heat for isospin-symmetric matter is
considered.
\end{abstract}


\maketitle

\section{Introduction}

The in-medium nuclear interaction gives rise to a complex
thermodynamical phase diagram. Several phases are expected to take
place as temperature and density are varied, as for instance
superfluid, superconducting, boson condensed, Bose-Einstein
condensed deuterons, and quark deconfined phases. Along these
thermodynamical changes some constraints must be fulfilled, such as
conservation of baryonic number, electric charge, etc., which have
severe consequences on the evolution of the state of matter.\\
As a typical situation we can mention the liquid-gas phase
transition (LGPT), expected to occur in the low density-low
temperature regime of the nuclear matter. It is a subject of study
since long time ago, and it has received renewed attention in
recent years \cite{SHARMA-PAL,RIOS,CARBONE,WU,HEMPEL}. The
theoretical treatment requires an interesting combination of
quantum statistical approaches with models of the nuclear
interaction. The contrast with empirical results could be made in
the field of ion collision experiences, see for example
\cite{HUANG-BONASERA}, as well as with observational data
concerning the thermal equilibration of proto-neutron stars
\cite{YAKOVLEV, GNEDIN}. Different situations prevent a direct
application of the theoretical predictions, for instance, finite
size effects are significant in heavy ion collisions. Furthermore,
the very short characteristic times of the reactions, turn it
dubious the applicability of equilibrium thermodynamics.\\
On the other hand, the possible frustration of the binodal
transition and the coming up of a non-homogeneous phase near the
surface layer of neutron stars complicate the interpretation of
the transport properties of the star matter.\\
It is clear that detailed calculations should take into account a
multitude of specific effects. However, with the purpose of
highlight the basic features, some simplified calculations are
admissible and they serve as a useful reference for more complex
statements \cite{HEMPEL}.\\
The existence of more than one conserved charge is a feature of
the above mentioned situations. A phase transition taking place
under such requirements has distinctive consequences, such as the
fact that the conserved charges do not distribute uniformly among
the coexisting phases \cite{MULLER-SEROT}. Certainly, the isospin
fractionation observed in multifragmentation experiences
\cite{RANDRUP} could be the fingerprint of a LGPT occurring after
the collision.

A variety of models and approximations have been used to describe
the nuclear equation of state. The combination of mean field
approaches with effective models, adjusted to reproduce the
nuclear phenomenology, offers the advantages of simple
calculations and reliable results. Among the most used
representations of the nuclear force, we can mention the
non-relativistic Skyrme model and the covariant formulation known
as Quantum Hadro-Dynamics (QHD). They have dissimilar foundations,
a density dependent nucleon-nucleon potential and a covariant
exchange of mesons are respectively used, but eventually both give
rise to an energy density functional. Within
this formulation, they can be fairly compared.\\
The fact that nuclear systems could have more than one conserved
charge, has noticeable consequences on the evolution of the
thermodynamical instabilities. Indeed, there is a change in the
order of the transition, allowing a continuous variation of the
thermodynamic potentials. Discontinuities are relegated to the
first derivatives, corresponding, for instance, to the
compressibility and heat capacity of the system.\\
It is worthwhile to mention that the inclusion of other effects,
such as the Coulomb and surface tension forces, could change
dramatically this description \cite{MEKJIAN3}.\\
In the present work we intend to study the behavior of the heat
capacity of infinite nuclear matter within the coexistence region.
The heat capacity is of great significance, for instance, in the
evaluation of the rate of change of temperature through the outer
shell of young neutron stars \cite{GNEDIN}. It has also been
studied in relation to the nuclear multifragmentation, where it is
considered as an indicator of the LGPT \cite{MEKJIAN1,MEKJIAN2}. A
statistical model of multinucleon clusters was frequently used for
this purpose, focusing the calculations in the low isospin
asymmetry regime. The upper limit for the temperature is
determined by a characteristic value, of the order $T\sim 10$ MeV,
for which clusters start to dissolve.\\
 In this work we explore a
wide range of temperature, particle density and isospin
composition, which can be easily combined in a mean field
calculation. We examine two effective models of the nuclear
interaction: the Skyrme potential and the QHD relativistic formulation.\\
This article is organized as follows. The general features of the
models are presented in the next section, the results are shown
and discussed in Section \ref{Results}. A final summary is given
in Section \ref{Summary}.

\section{The nuclear liquid-gas phase transition in different models}\label{SecII}

In order to check the generality of the results found, two
phenomenological models of the nuclear force have been used. In
first place the well known Skyrme model, where medium effects are
included through density dependent parameters for the
nucleon-nucleon potential. As an alternative formulation, we
choose the QHD model. Here the interaction is mediated by meson
fields, which are evaluated in a self-consistent
way. \\
In both cases the isospin composition can be easily handled. It
can be parameterized by the asymmetry fraction $w=(n_2-n_1)/n$,
with $n_1, \, n_2$ standing for the particle number density of
protons and neutrons respectively, and $n=n_1+n_2$ is the total
nucleon density.\\
We assume both proton and neutron numbers are conserved
independently. Hence, different chemical potentials $\mu_a$ can be
assigned to each isospin component.  The statistical distribution
function can be written $f_a(T,p)=\left[1+\exp \beta
(\varepsilon_a(p)-\mu_a) \right]^{-1} $, where the particle energy
spectrum $\varepsilon_a(p)$ is provided by the proposed model.\\
 Throughout this article we use units such that
$c=1$, $\hbar =
1$, $k_B=1$.\\

\subsection{The Skyrme model} \label{Skyrme}

 The Skyrme model is a well known effective formulation of
the nuclear interaction \cite{BENDER1}. It consists of a basic
Hamiltonian with contact potentials and density dependent
coefficients
\begin{eqnarray}
v_{Sky}(r_1,r_2)&=&t_0 (1+x_0 \, P_\sigma)
\delta(r_1-r_2)+\frac{1}{2}t_1 (1+x_1 \, P_\sigma)
\left[\stackrel{\leftarrow}{q}^2
\delta(r_1-r_2)+\stackrel{\rightarrow}{q}^2
\delta(r_1-r_2)\right] \nonumber\\
&+&t_2 (1+x_2 \, P_\sigma)\stackrel{\leftarrow}{q} \cdot
\delta(r_1-r_2) \stackrel{\rightarrow}{q}+\frac{1}{6}t_3 (1+x_3 \,
P_\sigma) \delta(r_1-r_2) \rho^\gamma((r_1+r_2)/2) \nonumber \\
&+& i W_0 (\sigma_1+\sigma_2)\cdot \stackrel{\leftarrow}{q} \times
\delta(r_1-r_2) \stackrel{\rightarrow}{q}\nonumber
\end{eqnarray}
where $\sigma_k$ represent the Pauli matrices for spin,
$P_\sigma=(1+\sigma_1 \cdot \sigma_2)/2$ is the spin exchange
operator, and $q=-i (\nabla_1 - \nabla_2)/2$ is the relative
momentum operator. Several parameterizations have been given,
according to the applications planned. They cover from exotic
nuclei to stellar matter.\\
By taking the Hartree-Fock expectation value from this force, an
energy density functional is obtained, which for infinite
homogeneous nuclear matter is given by

\begin{equation}
 {\mathcal E}_{Skm}=\delta_s \sum_{j}\, \frac{K_j}{2
m^*_j}+\frac{1}{16}(a_0+a_2 w^2)\; n^2 \label{Enerdens}.
\end{equation}
the factor $\delta_s=2$ takes account of the spin degeneracy, and
the kinetic density of particles with isospin $j$ ($j$=1,2 for
protons and neutrons, respectively) is given by
\begin{equation}
K_j=\frac{1}{(2 \pi)^3}\int d^3p \;p^2 f_j(T,p) \label{Kindens}
\end{equation}
Here, $f_j(T,p)$ is the Fermi occupation number for the isospin
component $j$ at temperature $T$. The effective nucleon mass
$m^*_j$ for this state is given by
\begin{equation}
\frac{1}{m^*_j}=\frac{1}{m}+\frac{1}{4}\, n\,(b_0-b_2 w I_j)
 \label{EffMass}
\end{equation}
$m$ represents the in vacuum degenerate nucleon mass, and
$I_j=(-1)^{1+j}$. \\
The density dependent coefficients $a_0, \,a_2$ and $b_0, \, b_2$
can be expressed in terms of the standard parameters of the Skyrme
model
\begin{eqnarray}
a_0= 6 t_0 + t_3 n^\gamma, & b_0=[3 t_1 + t_2 (5 + 4 x_2)]/2 \nonumber \\
a_2= -2 t_0 (1 + 2 x_0)- t_3 (1 + 2 x_3) n^\gamma/3,& b_2= [t_2 (1
+ 2 x_2) - t_1(1 + 2 x_1)]/2\nonumber
\end{eqnarray}

Within  a Landau-Fermi liquid scheme, the particle spectra can be
obtained by the functional derivatives $\varepsilon_{as}(p)=\delta
\mathcal{E}_{Skm}/\delta f_{as}(T,p)$. In such a way, the
following result is obtained \cite{BAYM-PETHICK}
\begin{eqnarray}
\varepsilon_j(p)&=&\frac{p^2}{2 m^*_j}+\frac{1}{8} v_j + \Delta \varepsilon, \label{SPXpectrum} \nonumber\\
v_j&=&(a_0 - a_2 w I_j)\, n + \delta_s \sum_k (b_0 +I_j I_k b_2)
K_k \nonumber\\
\Delta \varepsilon&=&\left[3 n^2 -(1+2 x_3) n^2 w^2 \right]\sigma
t_3 n^{\sigma-1}/48 \nonumber
\end{eqnarray}

The set of self-consistent equations is completed with the
relation between the conserved particle numbers and the
corresponding chemical potentials
\begin{equation}
n_j= \frac{\delta_s}{(2 \pi)^3}\int d^3p \; f_j(T,p)
\label{Numdens}
\end{equation}

\subsection{QHD model}
This is a model of the covariant field theory, proposed to deal with
in-medium nuclear properties \cite{WALECKA}. The interaction is
mediated by iso-scalar mesons $\sigma$ and $\omega_\mu$, the first
one can be considered as a resonant state. In the present case the
isovector mesons $\phi$ and $\rho_\mu$ are also included, as well as
polynomial terms in the $\sigma$ field.  \\
The lagrangian density is
\begin{eqnarray}
{\cal L}&=& \bar{\Psi}\left(i \not \! \partial -M +g_s\, \sigma+g_c
\, \mathbf{\tau \cdot \phi}- g_w \not \! \omega - g_r \mathbf{\tau}
\cdot \not \! \mathbf{\rho}\right) \Psi + \frac{1}{2} (\partial^\mu
\sigma
\partial_\mu \sigma - m_s^2 \sigma^2)-\frac{A}{3}\, \sigma^3-\frac{B}{4} \,\sigma^4+ \nonumber
\\
&&\frac{1}{2} (\partial^\mu \phi \cdot \partial_\mu \phi - m_c^2
\phi^2) -\frac{1}{4}
 F^{\mu \nu} F_{\mu \nu} + \frac{1}{2} m_w^2
\omega^2 -\frac{1}{4}
 R^{\mu \nu}\cdot R_{\mu \nu} + \frac{1}{2} m_r^2
\rho^2\nonumber
\end{eqnarray}

\noindent where $\Psi (x)$ is the isospin multiplet nucleon field,
$F_{\mu \nu}=\partial_\mu \omega_\nu-\partial_\nu \omega_\mu, \;
R_{\mu \nu}=\partial_\mu \rho_\nu-\partial_\nu \rho_\mu$, and
$g_s, g_c, g_w, g_r, A,$ and $B$ are coupling constants. The
nonlinear self-interaction of the $\sigma$ meson is necessary to
obtain an adequate behavior for the incompressibility around the
saturation density.

Within a mean field approximation, the equations of motion are\\
\begin{eqnarray}
\left(i \not \! \partial -M +g_s \sigma+ g_c \tau_3 \phi- g_w
\gamma_0 \omega -g_r \gamma_0 \tau_3 \rho \right)
\Psi&=&0, \label{NUCLEONEQ} \\
m_s^2 \sigma + A \sigma^2 + B \sigma^3=g_s \sum_j n_{sj},
\;\;\;m_w^2 \omega& =& g_w \sum_j n_j \label{OMEGA} \nonumber\\
m_c^2 \phi = g_c \sum_j I_j \, n_{sj}\label{PHI}
\;\;\;m_r^2 \rho &=& g_r \sum_j I_j \, n_j \label{RHO} \nonumber.
\end{eqnarray}

As in the previous section, the density of particles with isospin
projection $j$ is represented by $n_j =<\bar{\Psi}_j \gamma_0
\Psi_j>$, whereas $n_{sj} =<\bar{\Psi}_j \Psi_j>$ was used for the
scalar density. They can be explicitly written as
\[
n_j= \delta_s \int \frac{d^3p}{(2 \pi)^3} f_j(p,T),\; \; \;
\;n_{sj}= \delta_s \int \frac{d^3p}{(2 \pi)^3} f_j(p,T)
\frac{m_j}{E_{pj}}\nonumber
 \]
where $m_j = m - g_s \, \sigma - g_c I_j \phi$ is the in-medium
effective mass and $E_{pj}=\sqrt{p^2+m_j^2}$. Due to the assumed
isotropy, only the zero component of the vector fields survives.
Furthermore as there are not decaying channels between nucleons,
only the third component of the iso-vectors contributes.\\
The statistical distribution function $f_j(p,T)$ depends on the
quasi-particle energies $\varepsilon_j= E_{pj}+ g_w \omega+ g_r
I_j \rho$.\\
The energy density can be obtained by first evaluating the
energy-momentum tensor and then taking mean values. In such a way,
it is obtained
\[
\mathcal{E}=\delta_s \sum_j \int \frac{d^3p}{(2 \pi)^3}
\,\varepsilon_j(p) f_j(p)+\frac{1}{2} m_s^2 \sigma^2 +\frac{1}{2}
m_c^2 \phi^2+\frac{1}{2} m_w^2 \omega^2+\frac{1}{2} m_r^2 \rho^2+
\frac{A}{3} \sigma^3+\frac{B}{4} \sigma^4
\]

\subsection{The low density nuclear phase transition}\label{secnum}

The entropy density for a system in thermodynamical equilibrium is
given by
\[
\mathcal{S}=-\delta_s \,\sum_j \int \frac{d^3p}{(2
\pi)^3}\,\left[f_j \log f_j+\left(1- f_j\right) \log \left(1-
f_j\right)\right]
\]

It can be used, together with the corresponding energy density
$\mathcal{E}$, to evaluate the free energy density
$\mathcal{F}=\mathcal{E}- T \mathcal{S}$ and the pressure
$P=\sum_j \mu_j \, n_j - \mathcal{F}$ of the system.\\
A homogeneous system at temperature $T$ and isospin composition
$n_1,\, n_2$ remains thermodynamically stable if the free energy
per unit volume $\cal{F}$ is lower than any linear combination of
energies corresponding to independent thermodynamical states, say
$a$ and $b$, satisfying the conservation laws \cite{MULLER-SEROT},
i. e.
\begin{equation}
\mathcal{F}(T,n_1,n_2)<\lambda \,
\mathcal{F}(T,n^{(a)}_1,n^{(a)}_2)+(1-\lambda)\,
\mathcal{F}(T,n^{(b)}_1,n^{(b)}_2) \label{STABILITY}
\end{equation}

Otherwise, the system becomes unstable and a change of phase is
feasible. In such a case, two states can coexist if they verify
the thermodynamical equilibrium conditions
\begin{eqnarray}
P(T,n^{(a)}_1,n^{(a)}_2)&=&P(T,n^{(b)}_1,n^{(b)}_2), \label{Gibbs1}\\
 \mu^{(a)}_1=\mu^{(b)}_1,&&  \mu^{(a)}_2=\mu^{(b)}_2  \label{Gibbs2}
\end{eqnarray}
 Within the coexistence region, the total density of particles is a
combination of contributions coming from each phase
\begin{equation}
n_k=\lambda \, n^{(a)}_k+(1-\lambda)\, n^{(b)}_k, \;0<\lambda<1,
\;\; k=1,2, \label{Global}
\end{equation}
 where the $\lambda$ parameter stands for the
partial volume fraction of the phase $a$. As a consequence, if the
system has global densities $n_1, n_2$,  within the binodal region
it could be composed of  phases with densities differing
considerably from the global values.\\
Any extensive thermodynamical quantity can be evaluated in a
similar way, for instance the free energy can be written
\cite{MULLER-SEROT},  $\mathcal{F}(T,n_1,n_2)=\lambda \,
\mathcal{F}(T,n^{(a)}_1,n^{(a)}_2)+(1-\lambda)\,
\mathcal{F}(T,n^{(b)}_1,n^{(b)}_2)$.

In practice, we proceed as follows. For a given temperature $T$,
we fix the pressure $P_0$ within a reasonable range. We explore
this isobar and find a set of values $(n^{(a)}_1,n^{(a)}_2),
(n^{(b)}_1,n^{(b)}_2)$ which fulfill Eq. (\ref{Gibbs2}). For these
pairs of states we find all the solutions $(n_1,n_2,\lambda)$
consistent with the requirement of Eq. (\ref{Global}). Exploring
the range of temperatures and pressures for which there exist
solutions of Eqs. (\ref{Gibbs1})-(\ref{Global}), we obtain the
binodal region immersed in a 3-dimensional space, say $(T,P,w)$.\\
An interesting situation occurs for certain values of the
variables $(T,P,w)$, for which the parameter $\lambda$ does not
exhaust the full range $[0,1]$. In such situations $\lambda=0$ at
low density, then it grows with the density, reaches a maximum
value and then comes back to zero. This phenomenon is known as the
retrograde transition \cite{MULLER-SEROT}, because the system
starts and ends at the same phase
but in between develops a germ of the other phase.\\
Within these prescriptions the thermodynamical potentials remain
continuous throughout the phase transition. Discontinuities are
relegated to their derivatives. Of special meaning are the heat
capacity, compressibility and thermal expansion coefficient. All
of them are evaluated at constant particle number, for instance
the heat capacity at constant volume is defined as
$c_v=\left(\partial
\mathcal{S}/\partial T \right)_{N_1,N_2,V}$.\\
Within the binodal this derivative must be done carefully, since the
total number of protons is distributed between the two coexisting
phases. Furthermore, the parameter $\lambda$ of Eq. (\ref{Global})
has also a temperature dependence which is not explicitly written.
Taken these facts into account, and using ${\mathcal E}_a={\cal
E}(T,n^{(a)}_1,n^{(a)}_2), \; {\mathcal E}_b={\cal
E}(T,n^{(b)}_1,n^{(b)}_2)$, the heat capacity per unit volume in the
binodal can be written
\begin{eqnarray}
c_v(T,n_1,n_2)&=&\lambda \, \left(\frac{\partial {\cal
E}_a}{\partial T}\right)_{n_1,n_2}+(1-\lambda)\left(\frac{\partial
{\cal E}_b}{\partial T}\right)_{n_1,n_2}+\left({\cal E}_a - {\cal
E}_b\right) \left(\frac{\partial
\lambda}{\partial T}\right)_{n_1,n_2} \label{Cv1}\\
\left(\frac{\partial {\cal E}_c}{\partial T}\right)_{n_1,n_2}&=&
R_c + \left(\frac{\partial {\cal E}_c}{\partial
T}\right)_{n^{(c)}_1,n^{(c)}_2} \label{Cv2} \\
R_c&=& \sum_{k=1,2} \left(\frac{\partial {\cal E}_c}{\partial
n^{(c)}_k}\right)_{T,n^{(c)}_j} \left(\frac{\partial
n^{(c)}_k}{\partial T}\right)_{n_1,n_2}, \;\; j\neq k \label{Cv3}
\end{eqnarray}
The last term in Eq. (\ref{Cv2}) can be recognized as the heat
capacity for a homogeneous system composed of only one phase
$c_v^{(c)}=c_v (T,n^{(c)}_1,n^{(c)}_2)$. Therefore, Eq.
(\ref{Cv1}) can be summarized as
\begin{eqnarray}
c_v(T,n_1,n_2)&=&\lambda c_v^{(a)}+(1-\lambda) c_v^{(b)}+ \Delta
c_v \label{Cv4}\\
\Delta c_v& =& \lambda R_a + (1-\lambda) R_b + \left({\cal E}_a -
{\cal E}_b\right) \left(\frac{\partial \lambda}{\partial
T}\right)_{n_1,n_2}  \label{Cv5}
\end{eqnarray}
When the system approaches to the binodal boundary from inside,
$\lambda \rightarrow 0$, or $\lambda \rightarrow 1$. For instance,
$c_v(T,n_1,n_2)   \rightarrow  c_v^{(a)}+R_a + \left({\cal E}_a -
{\cal E}_b\right) \left(\partial \lambda / \partial
T\right)_{n_1,n_2}$ when $\lambda \rightarrow 1$. Approaching the
same point, but from outside the binodal, yields $c_v(T,n_1,n_2)
\rightarrow  c_v^{(a)}$. Hence we have a discontinuity $R_a +
\left({\cal E}_a - {\cal E}_b\right) \left(\partial \lambda /
\partial T\right)_{n_1,n_2}$, where the
first term contains several derivatives evaluated at the binodal
boundary, while the second contribution is proportional to the
energy difference between the coexisting phases.
 Expressions for the several derivatives appearing in Eqs. (\ref{Cv1})-(\ref{Cv3}), can
be found in the Appendix.

\section{Results and discussion}\label{Results}

In this section we show and discuss the results obtained for the
binodal region and its thermodynamical properties as described by
the selected models of the nuclear interaction. In particular we
analyze the specific heat at constant volume throughout the phase
transition, and in the case of symmetric nuclear matter we consider
the definition of a latent heat and evaluate its
temperature dependence.\\
For the Skyrme model the SLy4 parametrization is used, for which
$t_0=- 2488.91$ MeV fm$^3$, $t_1= 486.82$ MeV fm$^5$, $t_2=-
546.39$ MeV fm$^5$, $t_3=13777$ MeV fm$^{7/2}$, $x_0=0.834, \;
x_1=-0.344, \; x_2=-1, \; x_3=1.354, \; \gamma=1/6$
\cite{DOUCHIN}.\\
For the QHD model with iso-vector mesons the parametrization given
by Ref. \cite{LIU} is used, for which  $(g_s/m_s)^2=10.33$ fm$^2$,
$(g_w/m_w)^2=5.42$ fm$^2$, $(g_c/m_c)^2=2.5$ fm$^2$,
$(g_r/m_r)^2=3.15$ fm$^2$, $A/g_s^3=0.033$ fm$^{-1}$,
$B/g_s^4=-0.0048$.\\
The saturation density, binding energy, incompressibility and
symmetry energy obtained are $n_0=0.159$ fm$^{-3}$, $E_B=-15.97$
MeV, $K=229.9$ MeV, $E_S=32$ MeV in the Skyrme model, and
$n_0=0.16$ fm$^{-3}$, $E_B=-16$ MeV, $K=240$ MeV, $E_S=30.5$ MeV
for the QHD model. Another significative quantity is the in-medium
nucleon mass $m^{\ast}$ at the saturation density, the values
$m^{\ast}/m=0.694$, and $m^{\ast}/m=0.75$ are
obtained for the Skyrme and QHD models, respectively.\\
In first place the binodal region is constructed for both models.
Some results corresponding to the temperatures $T=$5 and 10 MeV
are shown in Fig. 1, in a plot of pressure versus proton abundance
$y=(1- w)/2$. It can be seen that for the Skyrme interaction the
coexistence of phases extends up to larger pressures. The range of
temperatures, instead, is smaller. The critical temperatures are
14.5 MeV and 15.9 MeV for Skyrme and QHD, respectively. It is
worthwhile to mention that for high isospin asymmetries $w$ (low
$y$) the transition is of retrograde character. This situation can
be appreciated more clearly in a plot of the pressure as a
function of the global particle density, and several isospin
asymmetries, as shown for $T = 10$ MeV in Fig. 2. Continuous lines
correspond to the physical results, whereas dashed lines represent
non-equilibrium states previous to the Gibbs construction. For
high neutron excess ($w=0.6$) the retrograde transition differs
only slightly from the non-corrected pressure. In this
circumstance, the mechanical stability condition ($\partial
P/\partial n > 0$) is verified, but some of the matter diffusion
conditions $\partial \mu_1/ \partial n > 0, \;\partial \mu_2/
\partial n < 0$ is not fulfilled. Furthermore, from the same
figure it can be appreciated that within the non-equilibrium
region of highly asymmetric matter the Skyrme model presents
higher incompressibility than the QHD case. This is a consequence
of the fast increase of the pressure, but keeping almost unchanged
densities, chemical potentials, and specially the derivatives of
the chemical potentials. Hence, larger pressures are obtained in
the former case within the range of densities that do not satisfy
the equilibrium conditions. This fact explains why the binodal
region extends up to larger pressures in the Skyrme than in the
QHD calculations, as shown in Fig. 1. \\
The effects of the Gibbs construction on the free energy are shown
in Fig. 3. In order to ease the comparison, the rest mass
contribution has been removed from the QHD results. It can be
corroborated that the coexistence of phases effectively minimizes
the free energy, and changes its convexity also. For the
temperature shown, $T = 5$ MeV, there is a retrograde transition
for $w = 0.9$.\\
As a special case we consider matter globally iso-symmetric, in
such a case it behaves as a one component system
\cite{MULLER-SEROT}. Along the phase separation the coexisting
states have $w = 0$, and the pressure remains almost constant. In
this sense the LGPT resembles a first order transition. This
circumstance can be appreciated in Fig. 4, where the density
dependence of the pressure in the Skyrme model is shown for
several temperatures. The dashed line encloses the coexistence
area. As it was pointed out, there is no discontinuity in the
thermodynamical potentials and it is particularly true for the
entropy. Notwithstanding, the difference $ T \,
(\mathcal{S}(T,n^{(a)})-\mathcal{S}(T,n^{(b)}))$
 represents the heat transferred as the LGPT is accomplished
isothermically. It is interesting to compare this quantity with
the latent heat corresponding to a first order phase transition.
It must be noticed that some calculations find a first order LGPT
in the nuclear medium, even for two components systems. See for
instance \cite{MEKJIAN2}, where particle correlations beyond the
mean field approximation are included. Therefore the thermal
dependence of this variation could be used to characterize the
order of the
change of phases.\\
In a recent paper \cite{CARBONE} the latent heat for the LGPT in
symmetric nuclear matter was studied for several parameterizations
of the Skyrme model. In order to compare results, we consider the
quantity
\begin{equation}\label{Latent}L=T \,
(\mathcal{S}(T,n^{(a)})/n^{(a)}-\mathcal{S}(T,n^{(b)})/n^{(b)}),
\end{equation}
along an isothermal within the binodal, which coincides with the
definition of the specific latent heat for a first order
transition. For a given temperature, the coexisting states with
nucleon densities $n^{(a)}$ and $n^{(b)}$ are determined by the
conditions of equal chemical potentials and pressures, so that $L$
depends only on the temperature. In Fig. 5 the results obtained
for the SLy4 parametrization and nonlinear QHD model are shown.
The behavior is similar in both cases, with maximum values
$L_{max}=29.7$ MeV and $L_{max}=30.8$ MeV for Skyrme and QHD
forces, respectively. The value $L=0$ is reached at the critical
temperature. We corroborate some of the conclusions enunciated in
\cite{CARBONE}, $a$) when $T \rightarrow 0$, $L$ approaches to the
binding energy at the saturation density; $b$) for low
temperatures $L$ grows almost linearly, $c$) for the greater
$L_{max}$ corresponds the larger
critical temperature.\\
Now we consider the specific heat capacity at constant volume,
evaluated according to Eqs. (\ref{Cv1})-(\ref{Cv3}). In first
place we examine the results for the Skyrme model, as shown in
Fig. 6, where the density dependence of the heat capacity is
displayed for several isospin asymmetries. Dashed lines represent
the results obtained without the Gibbs construction. There are
discontinuities at the thresholds of the binodal, as discussed at
the end of section \ref{secnum}. For a given temperature, the
discontinuity decreases with the asymmetry $w$, as expected from
the fact that pure neutron matter does not exhibit instabilities
of the LGPT type. Within the binodal, $c_v$ is a decreasing
function of the density, in contrast to its behavior outside.
Comparing the upper and lower panels of this figure, a general
increment around 60\% is observed in the specific heat at $T =10$
MeV  respect to the $T=5$ MeV outcome. For all the curves shown,
there are jumps toward greater values of $c_v$ as the system reach
the pure liquid state. The only exception occurs for the greater
value of $w$ shown in each figure, for which a retrograde
transition takes place. The comparison with the results obtained
using the QHD model can be made by examining Fig. 7. There is a
general agreement with the previous description,  with slightly
greater values of $c_v$ corresponding to the QHD case. As a
particular situation, it can be mentioned that for $T=10$ MeV,
$w=0.6$ the curve for $c_v$ does not show an appreciable
discontinuity at the higher transition density $n=0.47\, n_0$,
because it is at the limit, in the isospin asymmetry variable,
separating full and retrograde
evolution.\\
The thermal dependence of $C_v$ is shown in Fig. 8, for some
selected values of the global density and asymmetry. For this
purpose we have chosen the Skyrme model, because the QHD results
do not differ qualitatively. The heat capacity per particle
exhibits an evident discontinuity of a few units at the threshold
of the binodal. An increasing behavior is obtained for the full
range of temperatures examined. As expected, this quantity
approaches to zero as the temperatures vanishes, according to the
Nernst principle \cite{CALLEN}. On the other hand for high enough
temperatures, the specific heat approaches asymptotically to the
non-interacting limit $3 k_B/2$. The degree of convergence to this
limit depends essentially on the global density $n$, with a
negligible influence of the isospin asymmetry. The transition
temperature is both a decreasing function of $w$ (for fixed $n$),
and $n$ (for fixed $w$). \\
It can be observed that the magnitude of the discontinuity
diminishes with both $n$ and $w$. For the lowest density shown
$n/n_0 = 0.2$ the greatest jump correspond to the lower isospin
asymmetry $w = 0.2$, at $T \simeq$ 14 MeV, and for the neutron
rich state $w = 0.8$ the discontinuity  at $T
\simeq$ 10.5 MeV decreases a 60 \%.\\
The results shown in Fig. 8(a) can be contrasted with Fig. 6 of
Ref. \cite{MEKJIAN2}. The comparison must be cautious since the
latter used a canonical description for a high (but finite) number
of particles in a inhomogeneous probe of nucleons. For A=1000 the
heat capacity increases softly with temperature, up to a
characteristic temperature $T \simeq$ 10 MeV, where reaches a
peaked maximum. For higher $T$ the heat capacity remains almost
constant. The difference $\Delta C_v/N$ between the maximum and
the plateau varies within the range 5-25, decreasing with the
asymmetry $w$. In contrast we obtain a high rate of growth before
the critical temperature, which is of the same order $T \simeq$ 10
MeV. Furthermore, the drop from peak to the plateau is lesser than
4. In our calculations the precise location of the temperature
corresponding to the maximum decreases with $w$, in opposition to
the behavior shown in \cite{MEKJIAN2}.

\section{Conclusions}\label{Summary}

In this work we have examined the behavior of the heat capacity of
infinite homogeneous nuclear matter in the region of low particle
density and low temperature, taking the isospin composition as a
relevant parameter. For this purpose we have selected two
different descriptions of the nuclear interaction. Both, the
non-relativistic Skyrme potential and the field theoretical QHD
model have been extensively used in the literature with remarkable
success. Although these effective models have very different
foundations, they describe appropriately the nuclear matter
phenomenology for sub-saturation densities. \\
We have examined the region of thermodynamical instability, where
nuclear matter separates spontaneously in different phases. As we
consider conservation of both neutron $n_2$ and proton $n_1$
densities, it is found a three dimensional region of coexistence
of phases. As a consequence the thermodynamical potentials are
continuous, leading to a second order phase transition. Under
these conditions,  the heat capacity at constant volume has been
particularly analyzed.\\
As a first approximation we neglected the electromagnetic
interaction, which could have significant influence on the
thermodynamical fluctuations leading to a change of phase, see for
example \cite{MEKJIAN3}.\\
The description obtained corresponds to a region of the phase
space  wider than in previous calculations, where more complex
states of matter were considered \cite{MEKJIAN1,MEKJIAN2}.\\
The Gibbs construction allows the conservation of the global
densities of each isospin component through the coexistence of two
phases with local densities differing appreciably from the global
values $n_1, n_2$. The relative abundance of each of these two
phases can be represented through a parameter $0 < \lambda < 1$.
The energy of the system is expressed as a linear combination of
the energy of each phase, with coefficients $\lambda$ and $1-
\lambda$. The evaluation of the heat capacity requires some care,
since its definition prescribes derivatives at constant global
densities, which does not imply fixed local densities. The
temperature dependence of $\lambda$ must also be considered.\\
We have found qualitative agreement between the predictions of
both models. Only small differences can be found, for instance in
the extension of the binodal region, the critical temperature and
the maximum value of the specific heat capacity.\\
As expected in a second order phase transition, the heat capacity
exhibits a discontinuity at the boundary of the binodal. A
detailed characterization of this discontinuity has been presented
in terms of the thermodynamical variables for the full range of
the coexistence of phases. For a fixed temperature we found a
discontinuity at very low density and other one at a relatively
greater value, corresponding to the passage to pure gas and pure
liquid, respectively. As the isospin asymmetry is increased, the
full transition is replaced by a retrograde one. The high density
discontinuity in $c_v$ has opposite behavior for each of these
situations. For instance, in the retrograde evolution, $c_v$
decreases suddenly when the matter leaves the coexistence region
towards the pure liquid phase.\\
The thermal variation of the specific heat, at fixed density or
isospin asymmetry, also shows a sharp but finite jump at a
characteristic value. The location of this critical temperature
decreases with both $n$ and $w$.\\
The magnitude of the discontinuity in the heat capacity per
particle is lesser than 4, diminishing for increasing density and
asymmetry.\\
As a special situation, we examined the change of phase for
symmetric nuclear matter, which develops at very low pressures. In
such a case we examined the entropy variation between the final
states of an isothermal process within the binodal, and we have
compared it with the latent heat $L$, defined for a first order
phase transition. As a function of temperature $L(T)$ has a
maximum value and vanishes for the critical temperature. We have
found small differences between Skyrme and QHD predictions, and a
general agreement with recently published
results \cite{CARBONE}.\\

\appendix*
\section{}

The derivatives appearing in Eqs. (\ref{Cv1}), (\ref{Cv2}) are
obtained as solutions of a set of algebraic linear equations.  We
start taking derivatives of Eqs. (\ref{Global}) keeping constants
$n_1,\, n_2$, for $k=1,2$. After rearranging terms we obtain
\begin{eqnarray}
0&=&(n_1^{(a)}-n_1^{(b)})\left[\lambda \frac{\partial
n_2^{(a)}}{\partial T} + (1-\lambda) \frac{\partial
n_2^{(b)}}{\partial T}\right]+ (n_2^{(a)}-n_2^{(b)})\left[\lambda
\frac{\partial n_1^{(a)}}{\partial T} + (1-\lambda) \frac{\partial
n_1^{(b)}}{\partial T}\right] \label{Sys1}\\
\frac{\partial \lambda}{\partial T}&=&\left[\lambda \frac{\partial
n_1^{(a)}}{\partial T} + (1-\lambda) \frac{\partial
n_1^{(b)}}{\partial T}\right]/(n_1^{(a)}-n_1^{(b)}) \label{Sys2}
\end{eqnarray}

In the next step, derivatives of Eqs. (\ref{Gibbs2}) are taken,
giving
\begin{equation}
\left(\frac{\partial \mu_j^{(a)}}{\partial
T}\right)_{n_1^{(a)}\!\!\!,\,n_2^{(a)}}+\sum_{k=1,2}
\left(\frac{\partial \mu_j^{(a)}}{\partial
n_k^{(a)}}\right)_{n_l^{(a)}\!\!\!,\,T} \frac{\partial
n_k^{(a)}}{\partial T}= \left(\frac{\partial \mu_j^{(b)}}{\partial
T}\right)_{n_1^{(b)}\!\!\!,\,n_2^{(b)}}+\sum_{k=1,2}
\left(\frac{\partial \mu_j^{(b)}}{\partial
n_k^{(b)}}\right)_{n_l^{(b)}\!\!\!,\,T} \frac{\partial
n_k^{(b)}}{\partial T} \label{Sys3}
\end{equation}
where $j=1,2$ and $l\neq k$. When there is no explicit statement,
partial derivatives respect to $T$ are evaluated holding the
global densities $n_1, n_2$ fixed.\\
Writing $P(T,n_1,n_2)=\sum_b \mu_b \, n_b -
\mathcal{F}(T,n_1,n_2)$ before derivating Eq. (\ref{Gibbs1}),
leads to
\begin{equation}
\sum_{j=1,\,2}\left(n_j^{(a)}-n_j^{(b)}\right)\left[\left(\frac{\partial
\mu_j^{(a)}}{\partial T}\right)_{n_1^{(a)}\!\!\!,\,n_2^{(a)}}+
\sum_{k=1,2}\left(\frac{\partial \mu_j^{(a)}}{\partial
n_k^{(a)}}\right)_{n_l^{(a)}\!\!\!,\,T} \frac{\partial
n_k^{(a)}}{\partial T}\right]=\mathcal{S}_b-\mathcal{S}_a
\label{Sys4}
\end{equation}
with $\mathcal{S}_c=\mathcal{S}(T,n_1^{(c)},n_2^{(c)})$.\\
Eqs. (\ref{Sys1})-(\ref{Sys4}) constitute a set of five equations
in the unknowns $\partial n_j^{(c)}/\partial T, \,j=1,\,2, \,
c=a,\,b$ and $\partial \lambda/\partial T$. Some of the
coefficients are model dependent, for instance in the Skyme model
we have
\begin{eqnarray}
\left(\frac{\partial \mu_j^{(a)}} {\partial
n_k^{(a)}}\right)_{n_l^{(a)}\!\!\!,\,T}&=& \frac{\partial
\tilde{\mu}_j^{(a)}} {\partial
n_k^{(a)}}+(a_0-I_j I_k a_2)/8-\frac{\beta}{(8\pi)^2} \sum_i (b_0+I_i I_j b_2)^2 H_{i3}^{(a)} \nonumber \\
&&+\frac{\gamma t_3 n_a^{\gamma}}{48}\left[3 (\gamma+3)+ 2 w_a
(1+2 x_3)(I_k-I_j)+(1+2 x_3) (1-\gamma) w_a^2\right]
\nonumber \\
&&+\frac{\beta}{8 \pi^2} \sum_i (b_0+I_i I_j b_2) H_{i2}^{(a)}
\frac{\partial \tilde{\mu}_i^{(a)}}{\partial n_l^{(a)}}, \;\; l\neq k \nonumber \\
\frac{\partial \tilde{\mu}_j^{(a)}}{\partial n_k^{(a)}}&=&\left[
4(I_j I_k-1)+\beta(b_2 I_j I_k -b_0) H_{j\,
2}^{(c)}\right]/8\,\beta\,H_{j\,1}^{(c)}\nonumber \\
\left(\frac{\partial \mu_j^{(a)}} {\partial
T}\right)_{n_1^{(a)}\!\!\!,\,n_2^{(a)}}&=&\beta\left[\mu_j^{(a)}+\frac{v_j^{(a)}}{8}+\frac{\gamma
t_3 }{48}n^{(a)\,(1+\gamma)}\left(3-(1+2 x_3)
w_a^2\right)-\frac{H_{j2}^{(a)}}{2 m_j^{(a)} H_{j 1}^{(a)}}\right]
\nonumber \\
&&+\left(\frac{\beta}{4 \pi}\right)^2\sum_k(b_0+I_j I_k
b_2)\left(H_{k 3}^{(a)}-\frac{H_{k 2}^{(a)\,2}}{H_{k 1}^{(a)}}
\right)/m_k^{(a)} \nonumber
\end{eqnarray}
where the effective mass $m_k$ and potential $v_k$ have been given
in Sec. \ref{Skyrme}, and
\[
H_{j k}^{(a)}=\frac{1}{\pi^2}\int_0^\infty dp \, p^{2k}
f_j(T,n_1^{(a)},n_2^{(a)})
\left[1-f_j(T,n_1^{(a)},n_2^{(a)})\right].
\]
Furthermore
\begin{eqnarray}
\left(\frac{\partial \mathcal{E}_a}{\partial
T}\right)_{n_1^{(a)}\!\!\!,\,n_2^{(a)}}&=&\left(\frac{\beta}{2\pi}\right)^2
\sum_j \left(H_{j 3}^{(a)}-\frac{H_{j 2}^{(a)\;2}}{H_{j 1}^{(a)}}
\right)/m_j^{(a)} \nonumber \\
\left(\frac{\partial \mathcal{E}_a} {\partial
n_k^{(a)}}\right)_{n_l^{(a)}\!\!\!,\,T}&=&
\nonumber\frac{n^{(a)}}{8}(a_0-I_k w^{(a)})+\frac{1}{4}\sum_j
(b_0+b_2 I_j I_k) K_j^{(a)}+\frac{\gamma
t_3}{48}n^{(a)\,(1+\gamma)}\left(3-(1+2 x_3) w_a^2\right)\nonumber
\\
&&-\frac{\beta}{8 \pi^2} \sum_j \left[(b_0+I_j I_k b_2) \left(
H_{j 3}^{(a)}+\frac{H_{j 2}^{(a)\; 2}}{H_{j 1}^{(a)}}
\right)\right]/m_j^{(a)}-\frac{1}{2} \sum_j (1+I_j  I_k)
\frac{H_{j 2}^{(a)}}{m_j^{(a)} H_{j 1}^{(a)}}\nonumber
\end{eqnarray}
Similar calculations have been carried out within the QHD model,
but in this case further complications arise because the
interaction is mediated by the meson fields.

\section*{Acknowledgements}This work was partially supported by the
CONICET, Argentina.

\newpage
\begin{figure}
\includegraphics[width=0.8\textwidth]{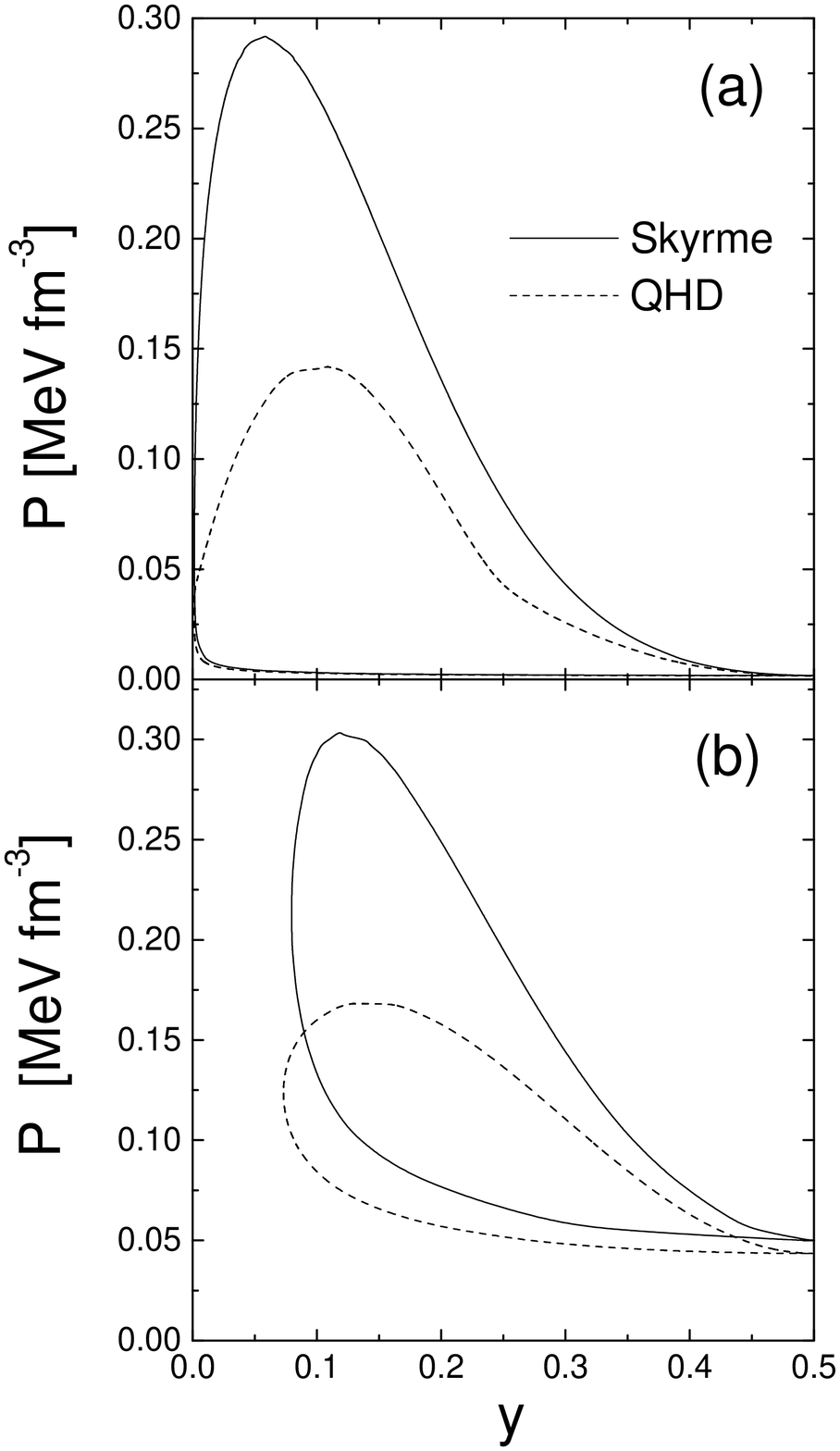}
\caption{\footnotesize Isothermal sections of the binodal
corresponding to T=5 MeV (a) and T=10 MeV (b), for the selected
models. The line convention specified in (a) is used for both
cases.}
\end{figure}

\newpage
\begin{figure}
\includegraphics[width=0.8\textwidth]{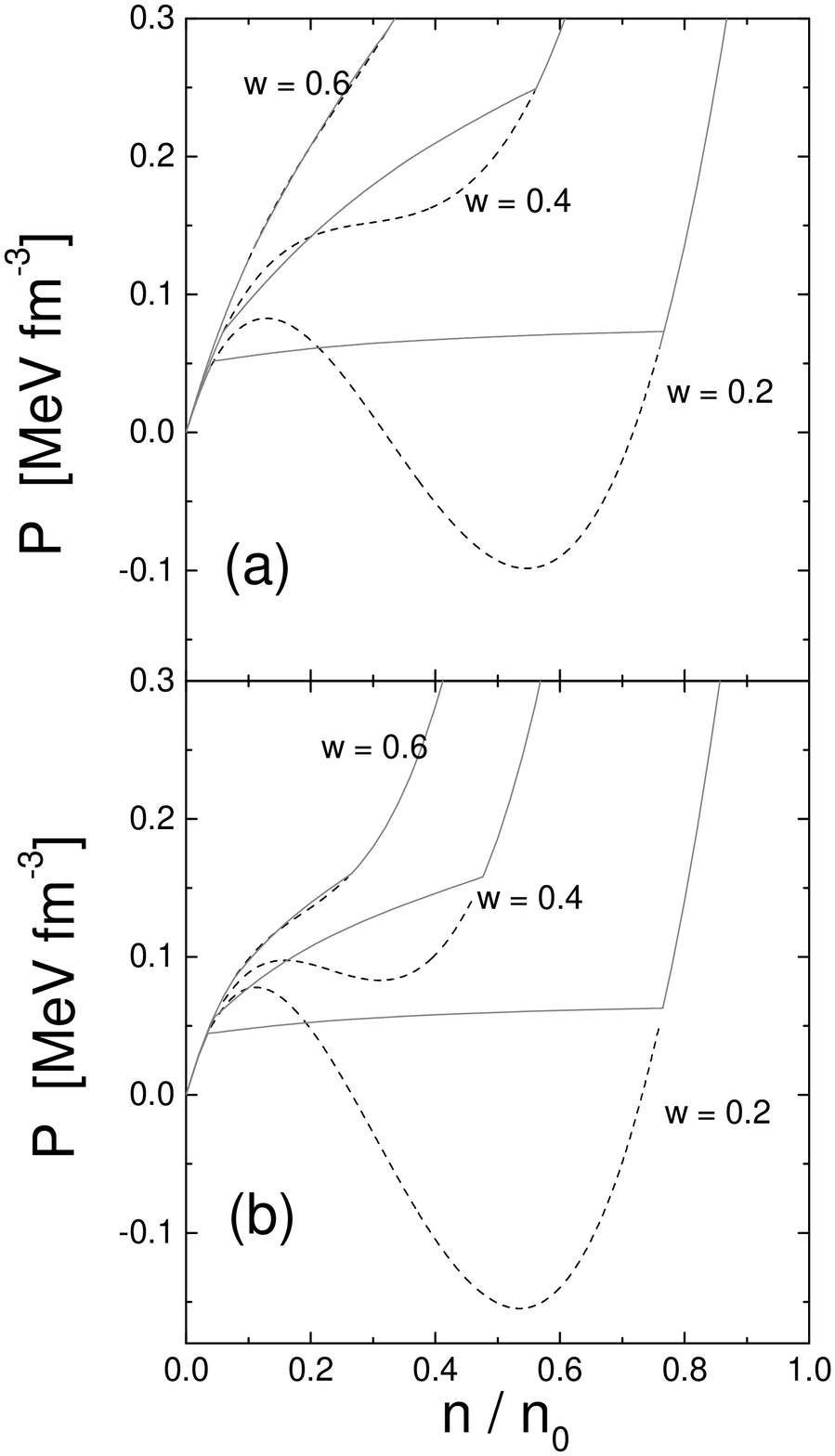}
\caption{\footnotesize The pressure as a function of the global
density of particles corresponding to T=10 MeV and several isospin
asymmetries w, for the  Skyrme (a) and QHD (b) models. Continuous
lines correspond to physical states in thermodynamical
equilibrium, dashed lines represent predictions of the models
without the Gibbs construction.}
\end{figure}
\newpage
\begin{figure}
\includegraphics[width=0.8\textwidth]{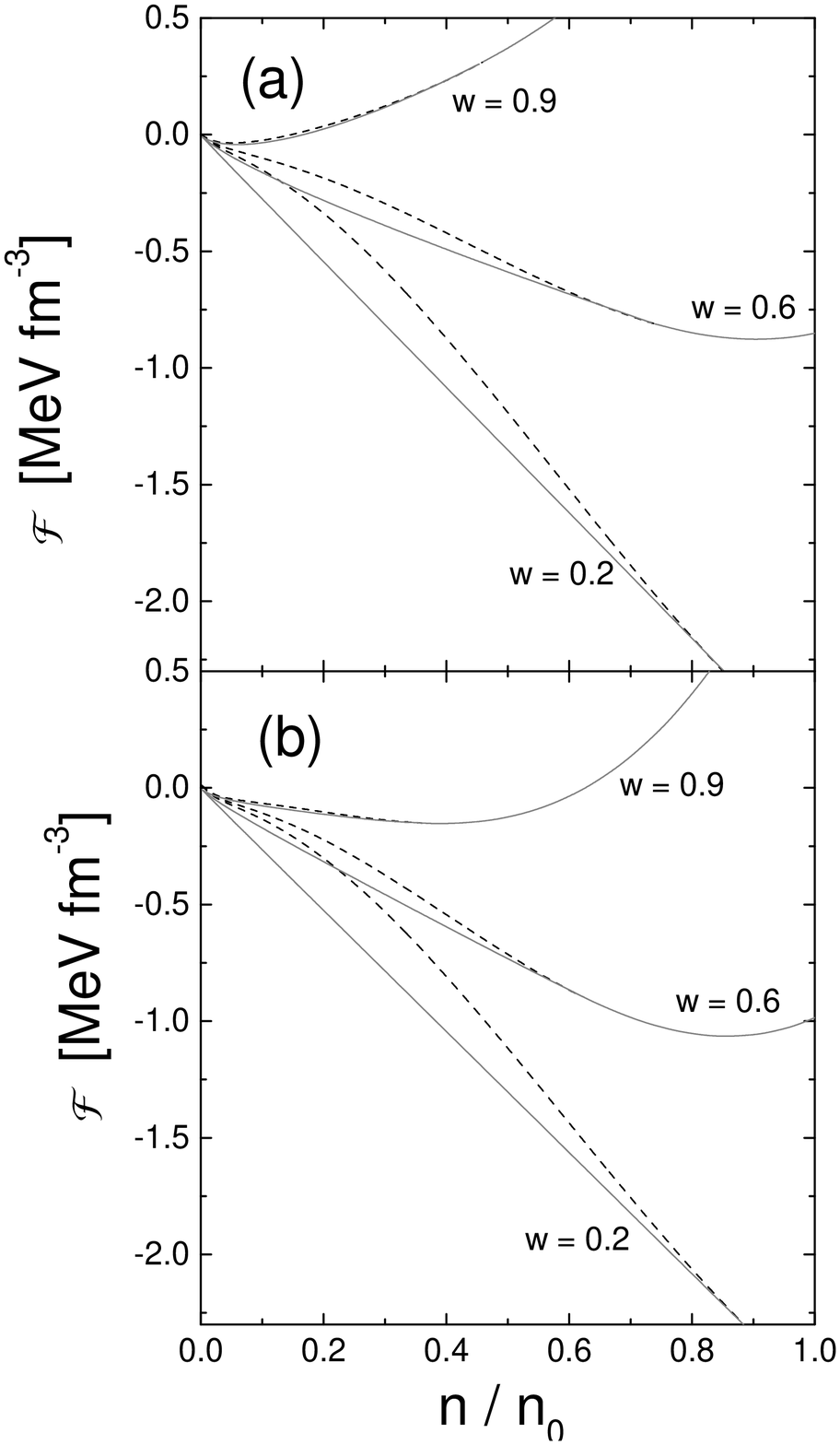}
\caption{\footnotesize The free energy density as a function of
the global density of particles corresponding to T=5 MeV and
several isospin asymmetries w, for the  Skyrme (a) and QHD (b)
models. Continuous lines correspond to physical states in
thermodynamical equilibrium, dashed lines represent predictions of
the models without the Gibbs construction.}
\end{figure}
\newpage
\begin{figure}
\includegraphics[width=0.8\textwidth]{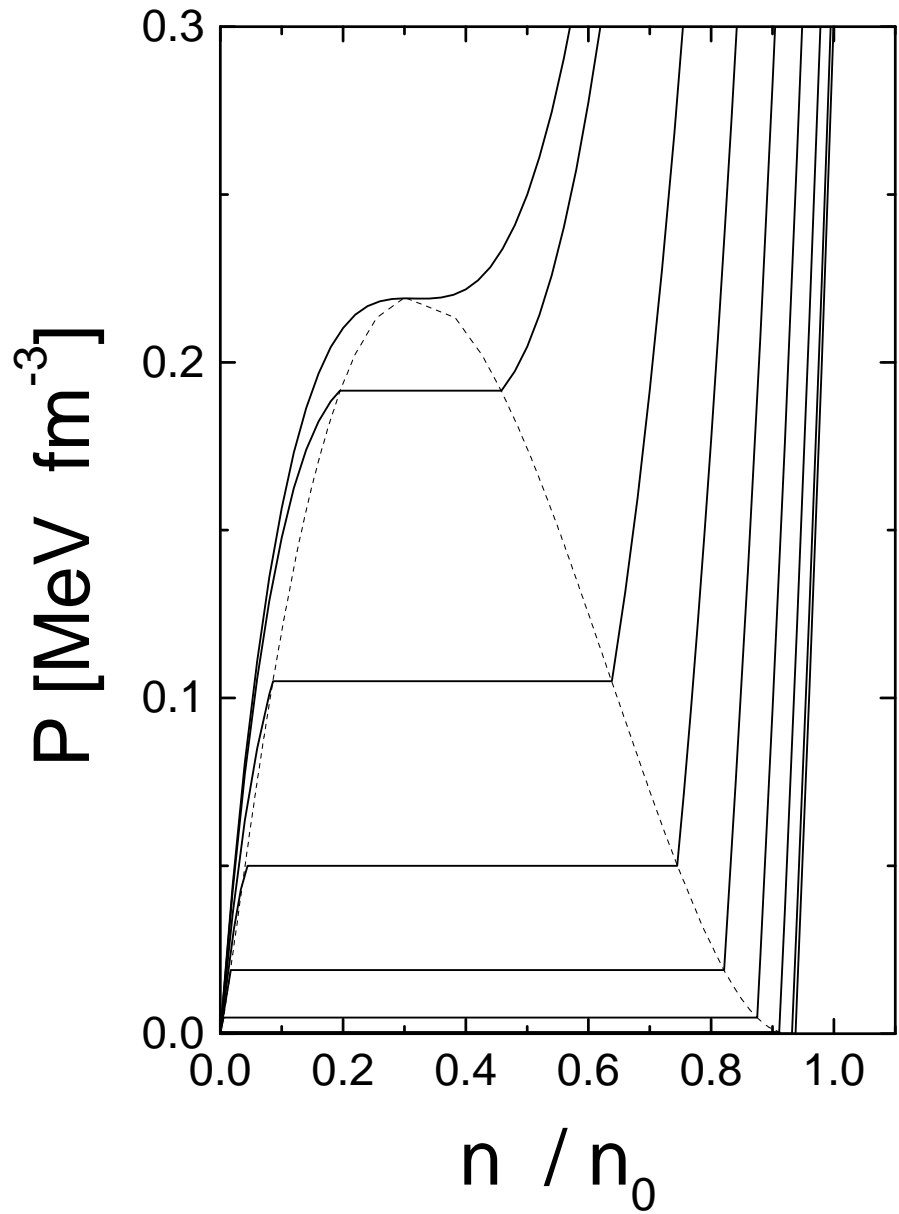}
\caption{\footnotesize The pressure for isospin symmetric nuclear
matter as a function of the global density of particles, for a set
of temperatures ranging from $T$ =0 to the critical temperature
$T_c$ = 14.5 MeV within the Skyrme model. The dashed line
represent the boundary of the binodal.}
\end{figure}
\newpage
\begin{figure}
\includegraphics[width=0.8\textwidth]{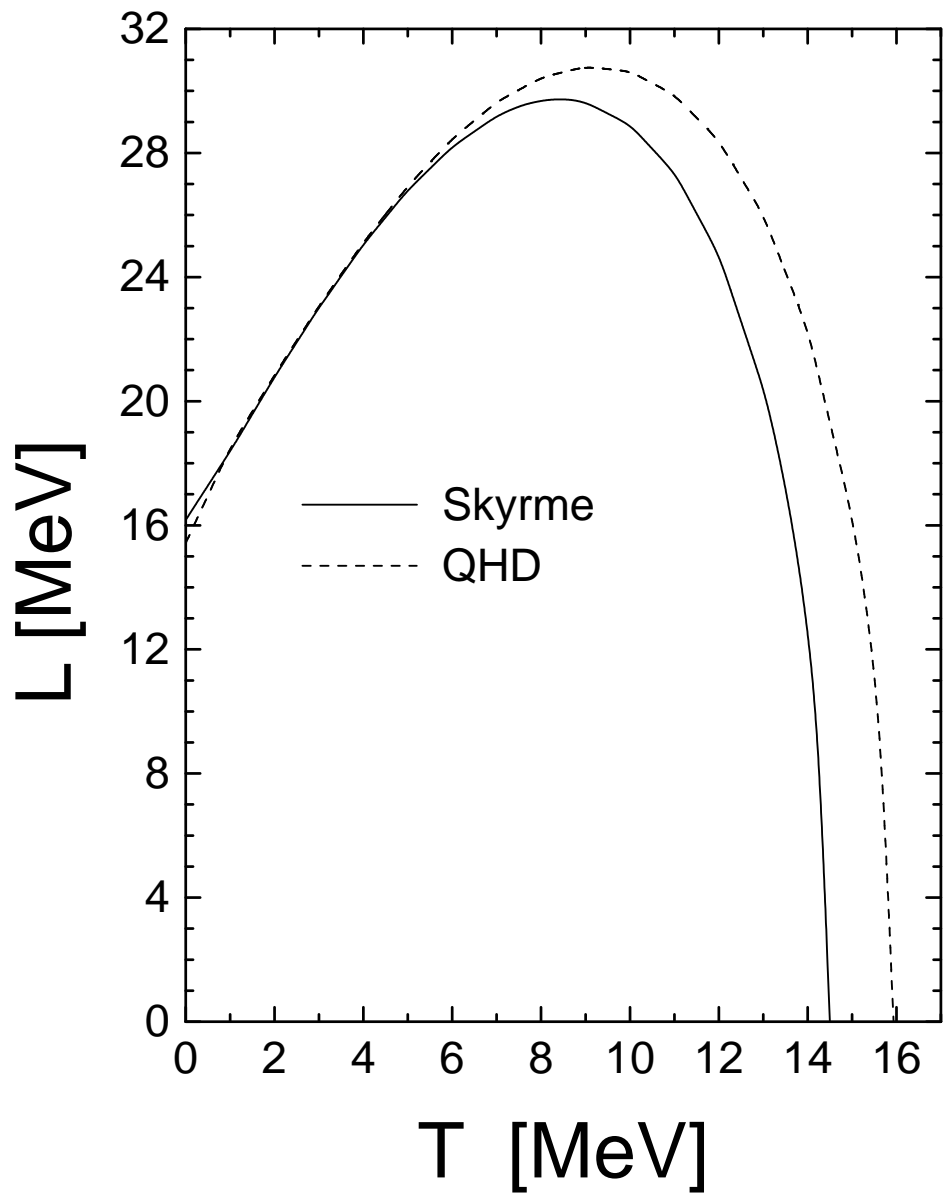}
\caption{\footnotesize The thermal dependence of $L$, see Eq.
\ref{Latent}, corresponding to the LGPT for symmetric nuclear
matter, as given by the selected models.}
\end{figure}
\newpage
\begin{figure}
\includegraphics[width=0.8\textwidth]{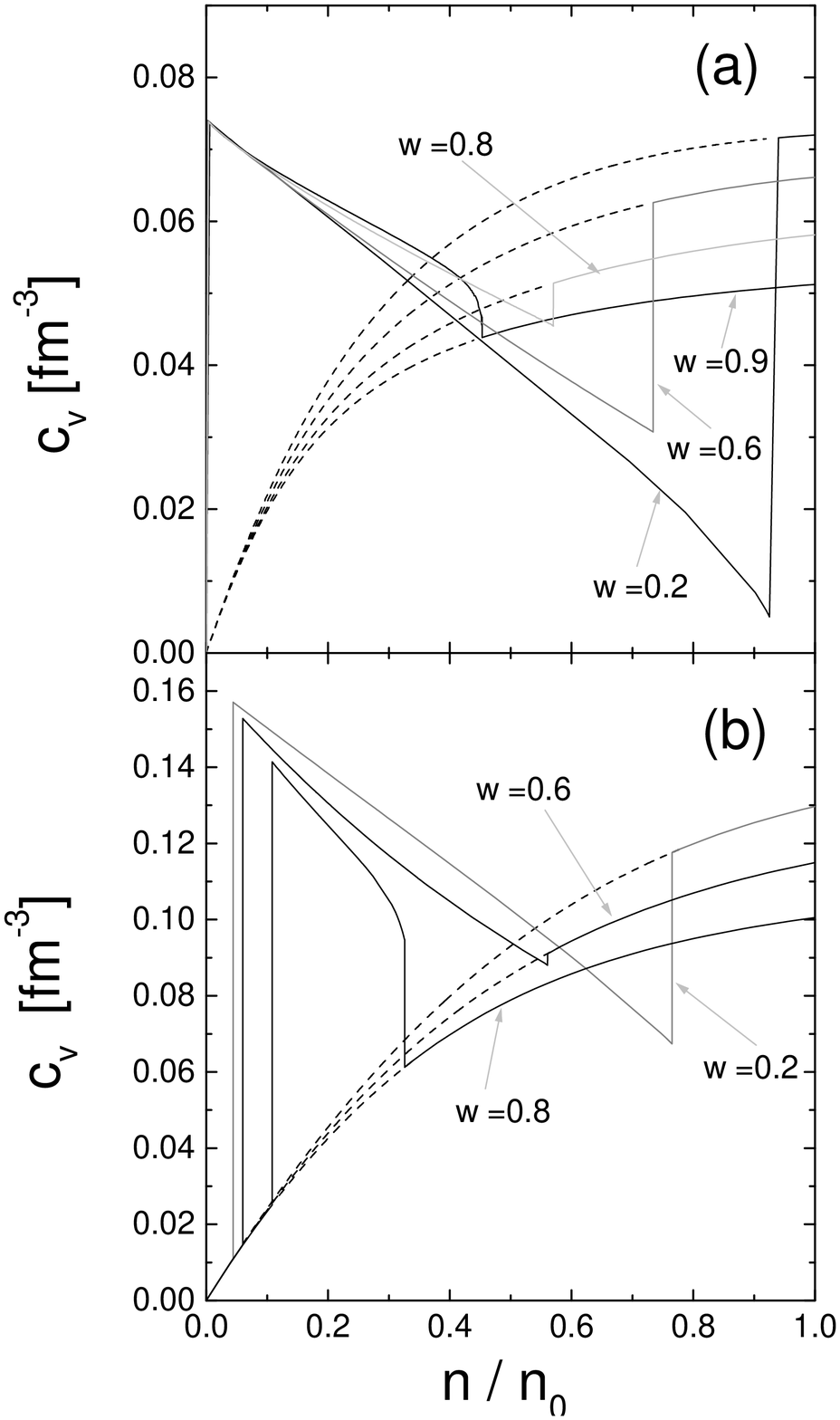}
\caption{\footnotesize The heat capacity per unit volume in terms
of the global density of particles for several asymmetries w,
corresponding to T=5 MeV (a) and T=10 MeV (b), within the Skyrme
model.  Continuous lines correspond to physical states in
thermodynamical equilibrium, dashed lines represent predictions of
the model without the Gibbs construction.}
\end{figure}
\newpage
\begin{figure}
\includegraphics[width=0.8\textwidth]{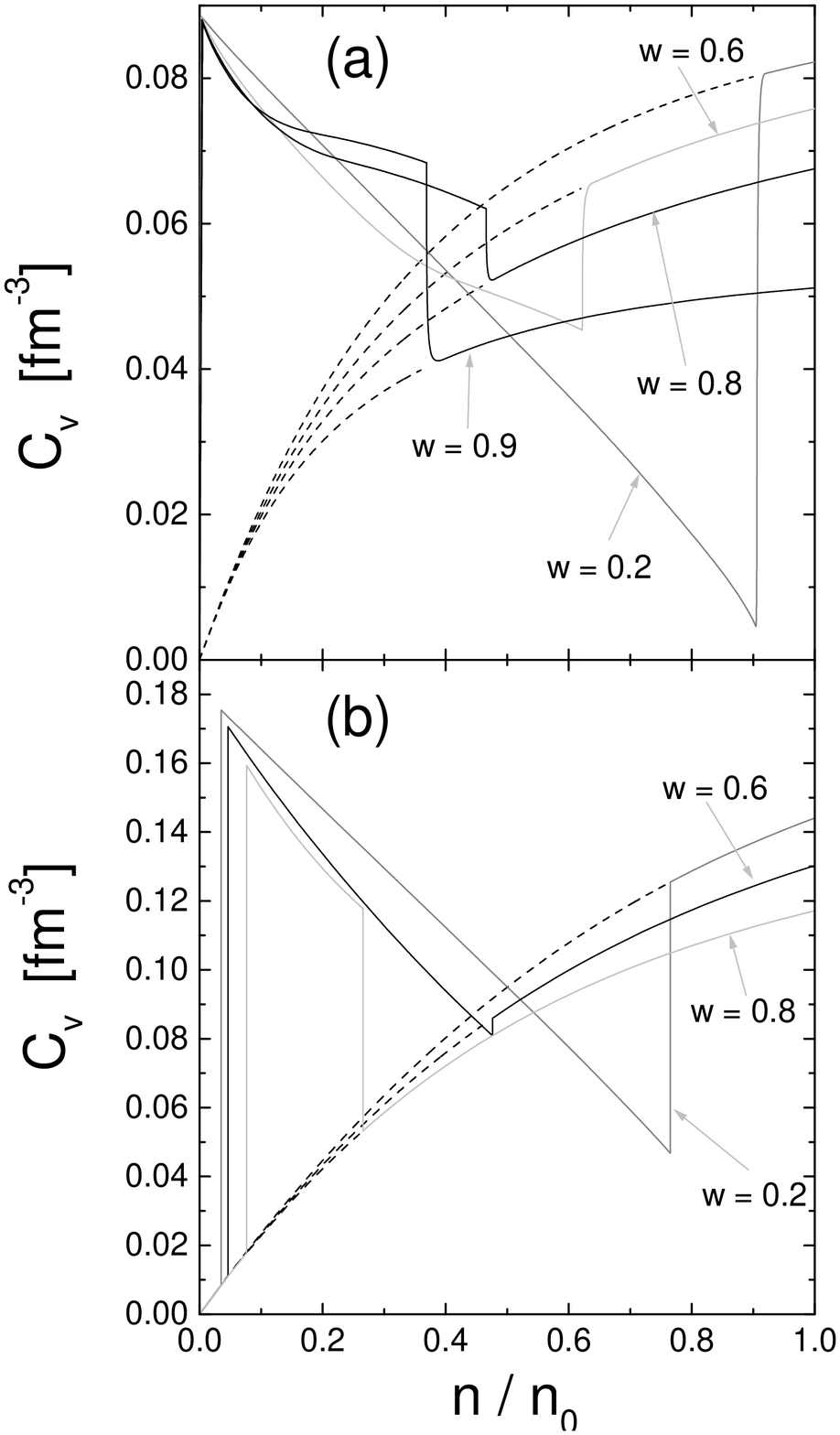}
\caption{\footnotesize The heat capacity per unit volume in terms
of the global density of particles for several asymmetries w,
corresponding to T=5 MeV (a) and T=10 MeV (b), within the QHD
model.  Continuous lines correspond to physical states in
thermodynamical equilibrium, dashed lines represent predictions of
the model without the Gibbs construction.}
\end{figure}
\newpage
\begin{figure}
\includegraphics[width=0.8\textwidth]{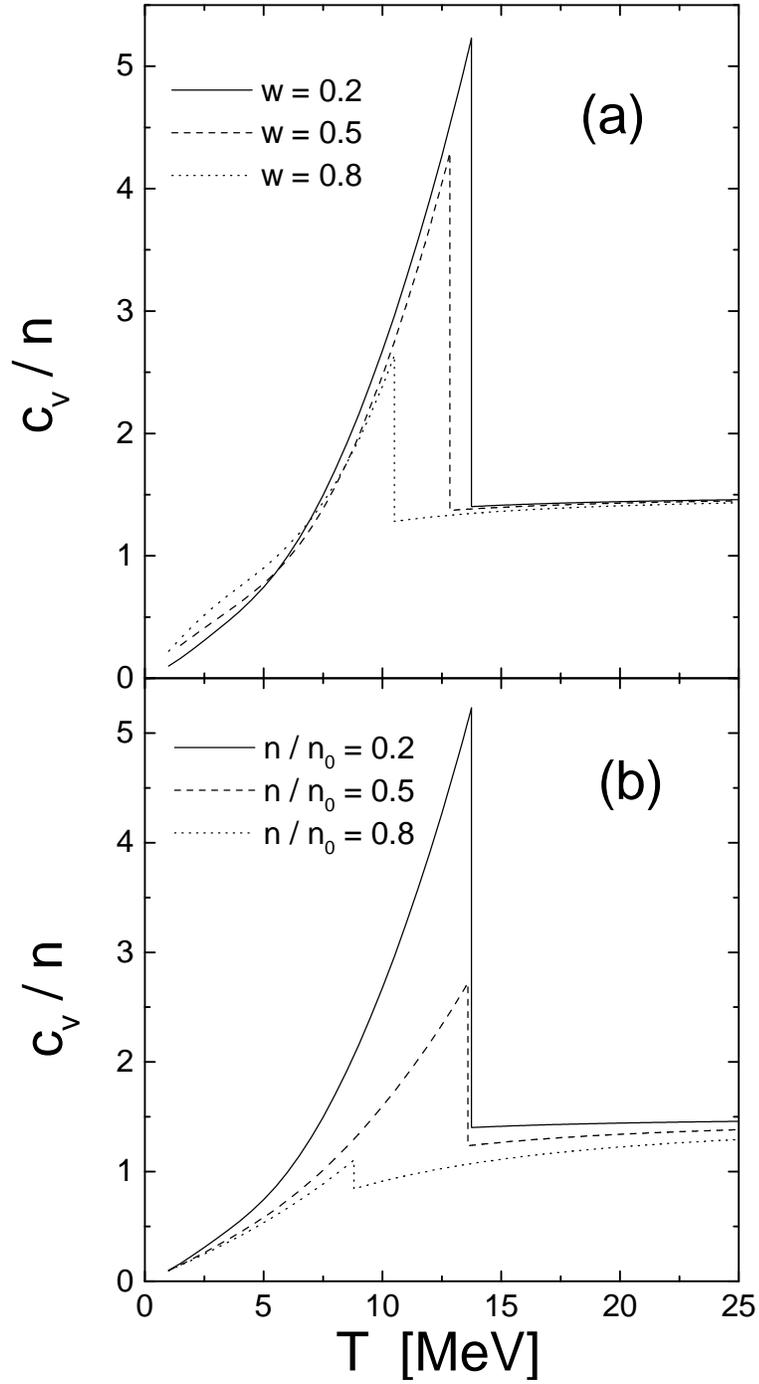}
\caption{\footnotesize The heat capacity per particle, within the
Skyrme model, in terms of the temperature for $n/n_0 = 0.2$ and
several isospin asymmetries w (a), and for fixed asymmetry w = 0.2
and several global densities (b).}
\end{figure}
\end{document}